\documentclass[aps, pra, reprint, superscriptaddress, preprintnumbers]{revtex4-1}

\usepackage{graphicx}
\usepackage{dcolumn}
\usepackage{bm}
\usepackage{braket}
\usepackage{color}
\usepackage{amsmath}

\begin{document}


\title{Experimental demonstration of two-photon magnetic resonances in a single-spin-system of a solid }



\author{T. Tashima}
\email[]{tashima.toshiyuki.5e@kyoto-u.ac.jp} 
\affiliation{These authors contributed to this work equally.}
\affiliation{Department of Electronic Science and Engineering, Kyoto University, 615-8510 Kyoto, Japan}

\author{H. Morishita}
\email[]{h-mori@scl.kyoto-u.ac.jp	}
\affiliation{These authors contributed to this work equally.}
\affiliation{Institute for Chemical Research, Kyoto University, 611-0011, Japan}


\author{N. Mizuochi}
\email[]{mizuochi@scl.kyoto-u.ac.jp}
\affiliation{Institute for Chemical Research, Kyoto University, 611-0011, Japan}


\date{\today}

\begin{abstract}
While the manipulation of quantum systems is significantly developed so far, achieving a single-source multi-use system for quantum-information processing and networks is still challenging.   A virtual state, a so-called ``dressed state," is a potential host for quantum hybridizations of quantum physical systems with various operational ranges. We present an experimental demonstration of a dressed state generated by two-photon magnetic resonances using a single spin in a single nitrogen-vacancy center in diamond. The two-photon magnetic resonances occur under the application of microwave and radio-frequency fields, with different operational ranges. The experimental results reveal the behavior of two-photon magnetic transitions in a single defect spin in a solid, thus presenting new potential quantum and semi-classical hybrid systems with different operational ranges using superconductivity and spintronics devices.
\end{abstract}


\maketitle

\section{Introduction}
Quantum hybridization is a key technology for quantum-information processing and networks like the quantum internet~\cite{KimbleNature2008}. Quantum hybridization is realized by combining the characteristics of photons, spins, and charges~\cite{KurizkiPANS2015}. For example, a photon can be used as a flying qubit for sharing quantum information, while spins and charges can be used as static qubits for quantum computation, quantum repeaters, or quantum memories.   Recently, quantum hybrid systems have been theoretically and experimentally demonstrated~\cite{KurizkiPANS2015, ToganNat10, ZhuNat2011,KuboPRL11,ZhuNCom14, DoucePRA15, LiPRL16, LeeArXiv16, MatsuPRA16}. Current quantum hybrid systems mainly consist of two different physical systems~\cite{ToganNat10, ZhuNat2011, KuboPRL11,AfzeliusNJP13, ZhuNCom14, DoucePRA15, LiPRL16, LeeArXiv16}. However, the quantum internet will require more complex quantum-information processing, such as the processing and storing of information while simultaneously updating the information in a quantum-information circuit and network. To realize more complex quantum-information processing, a dressed state~\cite{EITReview05} can be used as a host for hybridization between physical systems as well as for combining different operational ranges by tuning with a probe field and a pump field. Dressed states can also store quantum states.

\begin{figure}
	\includegraphics[width=8cm,clip]{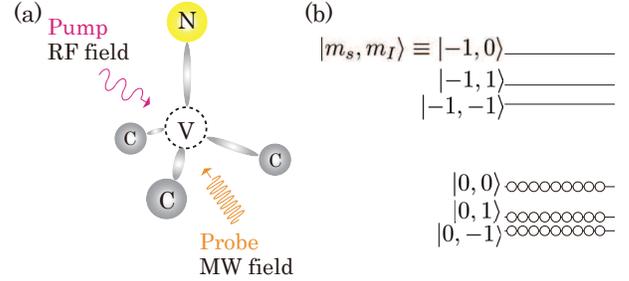}
	\caption{
		(color online) (a) Diagram of our experiment using two-photon magnetic resonance (b) Energy diagram of an NV center under a static magnetic field ($B_0$) along the NV-axis. 
	}
	\label{figSchematics}
\end{figure}

It is well known that nitrogen-vacancy (NV) centers in diamond are very useful for quantum-information processing, including entanglement generation~\cite{NeumannScience08}, quantum teleportation~\cite{PfaffScience2014}, quantum repeaters for linking nodes in networks~\cite{BernienNature2013}, quantum computation~\cite{WaldherrNature2014,TanimiuNatTech2014}, and implementing quantum memory~\cite{YanNPhoto2015}. NV centers in diamond are also one of the most promising candidates for quantum hybridization between different physical systems, since the quantum resources of photons, spins, and charges can be manipulated easily in NV centers. A quantum hybrid system between NV centers in diamond for accessing quantum-information networks and superconductors with good quantum computing performance, has been theoretically proposed and experimentally demonstrated using microwave (mw) operational fields~\cite{ZhuNat2011,KuboPRL11}.

To study the relevant fundamental behavior, a preliminary experimental generation of dressed states under optical and mw fields as a single operational range have been performed with ensembles of NV centers in diamond by electromagnetically induced transparency (EIT)~\cite{HePRB93, MansonJL10, AcostaPRL13, KehayiasPRB14, MrozekPRB16} and with single NV centers in diamond by coherent-population trapping~\cite{GolterPRL14,JamonneauPRL16}. However, these experiments were performed under the same operational ranges for both the pump and probe fields. Producing quantum technologies that combine physical systems of single qubits using different operational ranges remains a challenge. In our research, we have hybridized three or more physical systems of single qubits using different operational ranges, such as mw and radio-frequency (rf) fields. For example, NV centers can realize spins coupled in both mw and rf operational fields through the manipulation of the mw and rf operational fields. The key to realizing such systems is the generation of dressed states using multi-photon (here, two-photon) magnetic resonances on a single spin. A characteristic of this method is control of the resonant frequencies of the dressed states by tuning the mw and/or rf frequencies. The number of dressed states can be increased by adjusting the strength of the rf frequency via multi-photon magnetic resonance (discussed in Appendix C). Thus, the NV centers can hybridize three or more physical systems, so that a phenomenon requires further study. Recently, Childress et al. demonstrated multiphoton resonance in NV centers with mw and rf fields and discussed multiphoton resonances with the polarization of the $^{14}$N nuclear spin at $\sim$ 51 mT~\cite{ChildressPRA10}. Here, we experimentally generate a dressed state in a single NV center in diamond at 1.5 mT at room temperature by two-photon magnetic resonance (TPMR) via electromagnetic induction with mw and rf fields. We also demonstrate TPMRs below 5 MHz without polarization of $^{14}$N nuclear spins, and compare the experimental demonstrations with theoretical calculations in detail.

This paper is organized as follows: In Sec.~\ref{SecExp}, we briefly explain the experimental principle of the generation of dressed states. In Sec.~\ref{SecNV}, we explain our experimental setup consisting of a homemade confocal laser microscope with an irradiation system of mw and rf fields and our sample  .  In Sec.~\ref{SecEIVS}, we present preliminary experimental results for the NV center and discuss how the spins of the NV center are targeted. In Sec.~\ref{SecEst}, we present experimental results with two conditions for dressed-state generation using TPMR. In Sec.~\ref{SecDiscussion}, we estimate the characteristics of the generated dressed states by TPMR including the affect of the performance of our experimental system and the principle for the dressed-state generation of TPMR. Finally, in Sec.~\ref{SecConc}, we provide a brief summary and outlook for future research.

\begin{figure}
	\includegraphics[width=8.5cm]{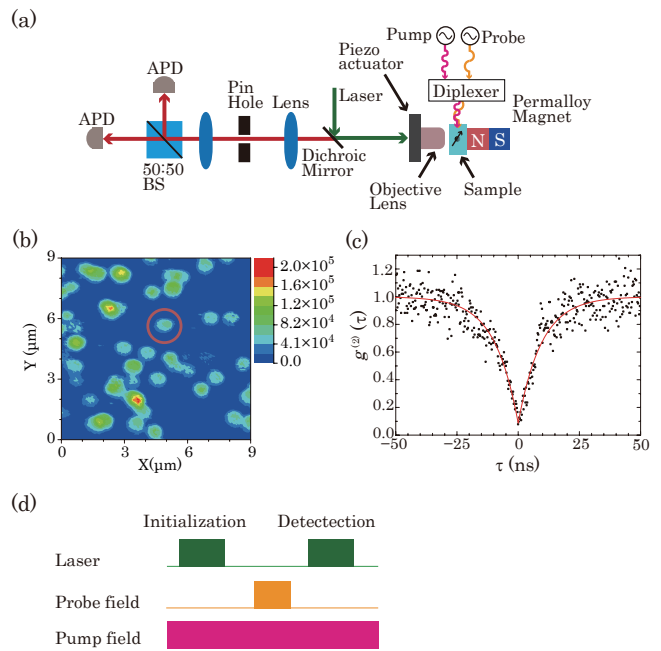}
	\caption{
		(color online) (a) Homemade confocal microscope with a system for electromagnetic field irradiation. A 532-nm laser excites an NV center. Photoluminescence is detected by two avalanche photodiodes (APD). Pump and probe fields are combined by a frequency diplexer and are irradiated on the sample via a thin copper wire. (b) Photoluminescence scanning-image of NV centers in diamond. The red circle shows the single NV center used in this experiment. (c) $g^{(2)}(\tau)$ of the fluorescence light emitted by the NV center. (d) Pulse sequence to generate a TPMR by means of a pump field. The pump field is irradiated continuously during both laser illumination and probe field irradiation.
	}
	\label{figSetup}
\end{figure}

\section{EXPERIMENTAL PRINCIPLE FOR THE GENERATION OF DRESSED STATES} \label{SecExp}
In this experimental demonstration, we used the electron spin and the $^{14}$N nuclear spin of the NV center to generate the dressed states. Figure~\ref{figSchematics}(b) shows the energy states of an NV center in diamond~\cite{HePRB93}. $\left|m_s, m_I\right>$ represents the electron spin and $^{14}$N nuclear spin of the NV center. After initialization of the electron spin by laser illumination, the population, depicted by the empty circles, is in the  $\ket{0, -1}$, $\ket{0,+1}$, or $\ket{0,0}$ states under a static magnetic field. To generate a dressed state, we used probe and pump fields near the electron and nuclear Zeeman energies. When the energy of the sum or difference of the pump and probe fields corresponds to the difference in the energy of the NV states, a dressed state is generated via TPMR for a single spin (see Appendix C). While three electron-spin resonance signals of the NV electron spin were observed in cases where dressed states were not generated, nine electron-spin resonance signals were observed when dressed states were generated. Namely, we can observe three signals from NV electron spins and six signals from the dressed states. During the dressed-state generation, even if the power of the pump field is changed, the generated dressed states behave in the same way. We measured the dependence of the TPMR frequencies of the dressed state generated by irradiation by the pump field, changing the pump power and detuning the probe frequency. We note that, in principle, the phase of the dressed states can be manipulated like an EIT~\cite{EITReview05}, but in this experiment, only population trapping with TPMRs is observed using a combination of continuous-wave rf fields and pulsed mw fields. Namely, we generate only the electromagnetically induced virtual structure.

\section{EXPERIMENTAL SETUP AND SAMPLE} \label{SecNV}
In our experimental setup, we used a homemade confocal laser microscope with an irradiation system of mw and rf fields, as illustrated in Fig.~\ref{figSetup}(a). A 532-nm laser, focused by an objective lens, was used to illuminate the NV center in the diamond sample. The NV center generated photoluminescence of 600 - 700 nm. The mw- and rf-field irradiation system was constructed on a sample stage. Two high-frequency oscillators created electromagnetic fields to manipulate the spins of the NV center. In our experiment, the mw field was around 2.8 GHz and the rf field was a few MHz. The mw and rf fields were combined by a frequency diplexer and were irradiated on the sample via a thin copper wire with a diameter of 10 $\mu$m. We used a high-temperature high-pressure (HTHP) type IIa (111) diamond (Sumitomo). To make single NV centers, $^{14}$N ions were implanted into the diamond at 500  $^{\circ}$C with a kinetic energy of 30 keV by a commercial ion implantation service, after which the sample was annealed at 750  $^{\circ}$C for 30 minutes.

\section{TARGET SPINS OF NV CENTER} \label{SecEIVS}
To identify a suitable NV center, we observed the photoluminescence scanning-image of a diamond with a laser at an optical wavelength of 532 nm, as shown in Fig.~\ref{figSetup}(b). The power of the laser was 100 $\mu$W. For the NV center marked by the red circle in Fig.~\ref{figSetup}(b), the second-order autocorrelation function, $g^{(2)}(\tau)$~\cite{BerPRL15}, was measured to be $g^{(2)}(0) \sim$ 0.1, shown in Fig.~\ref{figSetup}(c), which indicates a single NV center. We measured the optically detected magnetic-resonance (ODMR) spectra of the NV center with a pulsed laser (1-$\mu$s long) at a static magnetic field ($B_0$) of $\sim$ 1.5 mT by sweeping the probe frequency of the pulsed mw probe field (5.5-$\mu$s long). Figure~\ref{figPowerDepend}(a) shows the ODMR spectrum (brown plot), exhibiting three dips arising from the triplet hyperfine splitting of the $^{14}$N nuclear spin. We note that the pulsed 5.5 $\mu$s probe, which corresponds to a $\pi$ pulse, is weak enough to detect the hyperfine coupling between the electron spin and the $^{14}$N nuclear spin of the NV center.

\section{EXPERIMENTAL DEMONSTRATION OF TPMR} \label{SecEst}
We experimentally demonstrated the dressed-state generation via a TPMR. The TPMR was observed in the ODMR spectra by applying a continuous-wave rf pump field, as shown in Fig.~\ref{figSetup}(d), under two conditions. In the first condition, the pump power was changed while the pump frequency was fixed. In the second condition, the pump frequencies were detuned while the pump power was fixed.

\begin{figure}
	\includegraphics[width = 8cm, clip]{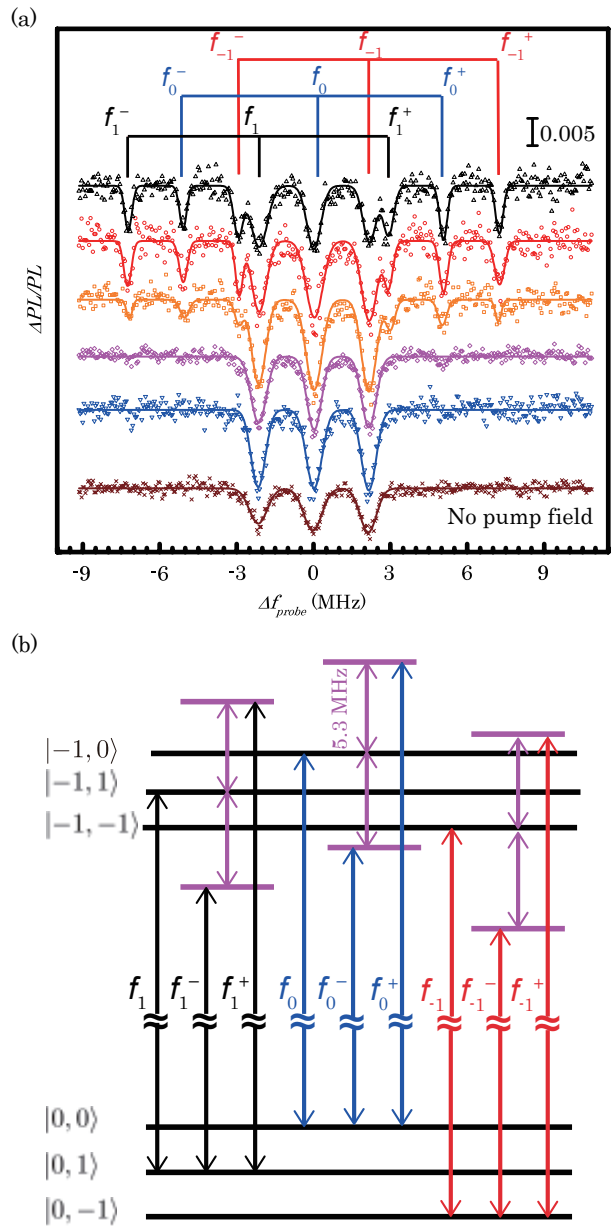}
	\caption{
	(color online) (a) ODMR spectra by sweeping the frequency of a 5.5-$\mu$sec probe field with/without irradiation by a continuous pump field, revealing the $^{14}$N nuclear hyperfine structure. The black, red, orange, pink, and blue plots show spectra for pump fields of 63.0 mW, 31.6 mW, 10.0 mW, 2.00 mW, and 0.63 mW, respectively. The brown plot shows the ODMR spectrum without a pump field. The solid lines show the fitting for each ODMR spectrum. The scale bar indicates a $\Delta PL/PL$ of 0.005. (b) Energy levels of a TPMR under the application of a pump field with a frequency of 5.3 MHz.
	}
	\label{figPowerDepend}
\end{figure}

In the first condition, we observed the dressed-states' positions by fixing the pump frequency at 5.3 MHz, while setting the pump power to 63.0 mW, 31.6 mW, 10.0 mW, 2.00 mW, and 0.63 mW. Note that a zero-detuning frequency corresponds to the center dips of the triplet hyperfine splitting of the $^{14}$N nuclear spin of the NV center. The results are shown in Fig.~\ref{figPowerDepend}(a). These experimental data can be fitted by three or nine Gaussian functions. When the pump power is less than 10 mW, the ODMR spectra have three dips, from the triplet hyperfine splitting of the $^{14}$N nuclear spin. When the pump power is over 10 mW, additional dips in the ODMR spectra appear without a change of the position of the original dips. Moreover, the hyperfine splitting of the $^{14}$N nuclear spin is observed under the application of the pump field, indicating that the hyperfine structure is not destroyed. Figure~\ref{figPowerDepend}(b) shows the theoretical energy levels of a TPMR under the application of a pump field of 5.3 MHz. Three of the nine signals correspond to the hyperfine splitting of the $^{14}$N nuclear spin. These three signals are indicated by $f_I$, where the subscript $I$ denotes the $^{14}$N nuclear spin. The other six signals are expected to appear when the following equation is satisfied: $f_I^\pm = f_I \pm f_\mathrm{pump}$, where $f_I^\pm$ is the resonance frequency of the TPMR and $f_\mathrm{pump}$ is the pump frequency, as illustrated in Fig.~\ref{figPowerDepend}(b). The solid lines in Fig.~\ref{figPowerDepend}(a) show the expected signal positions. The experimental results show excellent agreement with the theoretical predictions. Hence, it is clear that the dip positions do not depend on the power of the pump field. The linewidths at $f_I^\pm$ are narrower than those at $f_I$. For example, the linewidths at $f_1^-$ and $f_1$ under the application of a 63.0-mW pump field are 0.32 MHz and 0.71 MHz, respectively. Defining the dephasing time ($T_2^*$) as the inverse of the linewidth, the dephasing time of $f_1^-$ is 2.6 times longer than that of $f_1$. This result indicates that the influence of $^{13}$C nuclear spins on the dressed states becomes weaker under the application of the pump field. The coherence time of the dressed states can also be expected to become longer.

\begin{figure}
	\includegraphics[width=8.5cm,clip]{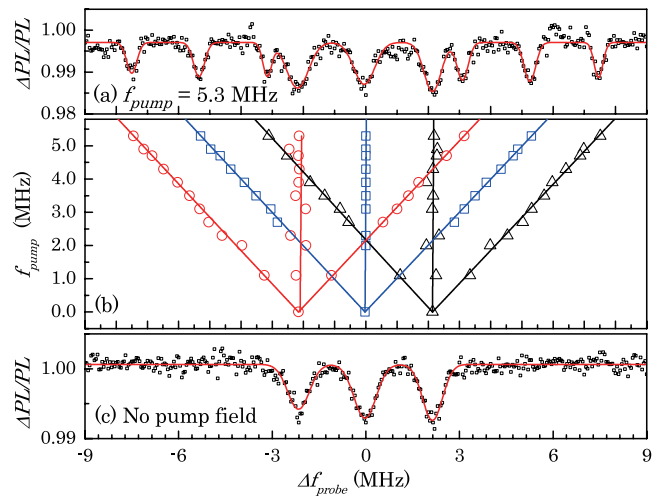}
	\caption{
	(color online) (a) ODMR spectrum with a 5.3-MHz pump field, as shown in Fig.~\ref{figPowerDepend}. The solid red line shows a fit using the sum of nine Gaussian peaks. (b) Peak position shifts due to changes in the pump field frequency. (c) ODMR spectrum without a pump field. The probe field has a length of 5.5 $\mu$sec and is sufficiently weak to allow the $^{14}$N nuclear hyperfine structure to be observed. The solid red line shows a fit using the sum of three Gaussian peaks.
	}
	\label{figFreqDepend}
\end{figure}

In the second condition, we observed the dip-position shifts by fixing the pump power while detuning the pump frequency. Figure~\ref{figFreqDepend}(a) shows the ODMR spectrum with a pump frequency of 5.3 MHz and a pump power of 63.0 mW. Figure ~\ref{figFreqDepend}(c) shows the ODMR spectrum without the pump field. Figure~\ref{figFreqDepend}(b) shows the location of the ODMR dips as a function of the pump frequency. The number of dips increases from three in Fig.~\ref{figFreqDepend}(c) to nine in Fig.~\ref{figFreqDepend}(a). The locations of the dips are extracted from the fit for each plot, and the spread of the dips is linear with respect to the pump frequency. Moreover, the absolute value of the splitting width corresponds well to the frequency of the pump field.

\section{ESTIMATION OF GENERATED DRESSED STATES BY TPMR} \label{SecDiscussion}
Here we estimate the characteristics of the dressed states produced with a ``diplexer" and a model for generating a TPMR. First, we consider the effect of the output frequencies of the diplexer on the ODMR spectra, although applying the pump field generates other energy levels. Figures~\ref{figSpeAna}(a) and \ref{figSpeAna}(b) in Appendix~\ref{AppDiplexer} show the output frequency of the diplexer as functions of pump power and pump frequency. They show that the diplexer output peaks satisfy the following conditions:$f = f_\mathrm{probe} \pm n \times f_\mathrm{pump}$, where $f$, $f_\mathrm{probe}$, and $f_\mathrm{pump}$ are as the peak frequency, probe frequency, and pump frequency, respectively, and $n$ is an integer. While the amplitudes of the frequencies in the diplexer output at $\left|n\right|$ = 2 are larger than those at $\left|n\right|$ = 1, the signals in this experiment were only observed with frequencies at $\left|n\right|$ = 1, as shown in Fig.~\ref{figPowerDepend}(a). These results imply that the effects of additional output frequencies from the diplexer are negligibly small. In addition, the fact that the length of a $\pi$-pulse depends on the probe power in our sequence for detecting TPMR also means that the effects are negligibly small. Second, we prove that the observed signals do not originate from nuclear-magnetic resonances in the NV center. Figure~\ref{figFreqDepend}(a) shows the nine dips that were observed in our experiments. The NMR signals of the NV center should appear at $f_0-Q-A$, $f_0-Q+A$, $f_1+Q-A$, and $f_{-1}+Q+A$, where $A$ is the hyperfine interaction between an electron spin and $^{14}$N nuclear spins in the NV center and $Q$ is the quadrupole interaction of the $^{14}$N nuclear spin (details discussed in Appendix~\ref{AppNV}). 
It is noted that we have neglected nuclear Zeeman energy because it is weaker than $A$ and $Q$ in these experimental conditions. Figure~\ref{figFreqDepend}(a) does not include any signals at the resonant frequencies of the NMR signals, namely, the effect of NMR is negligible small for our experimental conditions. Third, we examine the experimental results for the first conditions with the model for TPMR, as discussed in Appendix~\ref{AppMPNV}. The transitions described by Eq. (\ref{eqC8}) occur only under irradiation by a right-hand circularly polarized probe field and a linearly polarized rf pump field with respect to $B_0$. Equations (\ref{eqC8}) and (\ref{EqTPMRAmp}) show that the transition probability of the TPMRs depend on the strength of the pump field ($\omega_2$) (details discussed in Appendix~\ref{AppTPMR}) and the position depends on the pump frequency. Hence, additional dips appear when the sum or the difference between the pump and the probe frequencies corresponds to the triplet hyperfine splitting of the $^{14}$N nuclear spin, according to the model of Eq. (\ref{eqC8}). Also, transitions under two circularly polarized fields (e.g., $\Delta m_s$ = 2, $\Delta m_I$ = 0, $\Delta m_s$ = $\pm$1, and $\Delta m_I$ = $\pm$1) were not observed in this study. The results using TPMRs also show that electron magnetic resonances of the NV center occur when the NV center is dressed by the pump field. Therefore, the phenomena seen under the two experimental conditions demonstrate the generation of TPMR in the combined mw and rf regions. Also, the experimental results show that the quantum states due to TPMR are converted reversibly by both driving fields.

\section{Conclusion} \label{SecConc}
We now discuss the strategy for generating a hybrid system with a generated dressed state. L. Trifunovic et al.~\cite{TrifunovicPRX13} have proposed a hybrid strategy between magnetic materials and electron spin in NV centers in diamond using dipole coupling for a fusion of classical and quantum- information processing. In this case, our demonstration of TPMR indicates that we can hybridize these spins with wide-range operational fields through an NV center because we can optimize the parameters for coupling between both spins in both the mw and rf operational fields. Thus, our research of quantum hybridization with a generated dressed state is a promising candidate for this type of hybridization.
 
In conclusion, we have demonstrated TPMR using the single NV center in diamond at ambient conditions by driving the mw and rf fields without polarization of $^{14}$N nuclear spins. We clearly observed the generated dressed states using TPMRs via the single-defect spin in diamond. We showed that the generated dressed states can be converted reversibly by driving both fields. These results indicate the potential for hosting hybridizations of physical systems with different operational ranges. Thus, our results pave the way for a new fusion of physical systems with single qubits and wide operational ranges.

\section*{Acknowledgements}
We wish to thank M. Koashi and R. Inoue for their valuable comments in the early stages of the experiment. This work was supported by KAKENHI and CREST. TT was supported by a Grant-in-Aid for Young Scientists (B), Grant No. 15K17729.

\appendix{}
\section{Output frequency characterization of our diplexer} \label{AppDiplexer}
\begin{figure}
	\includegraphics[width=8.5cm,clip]{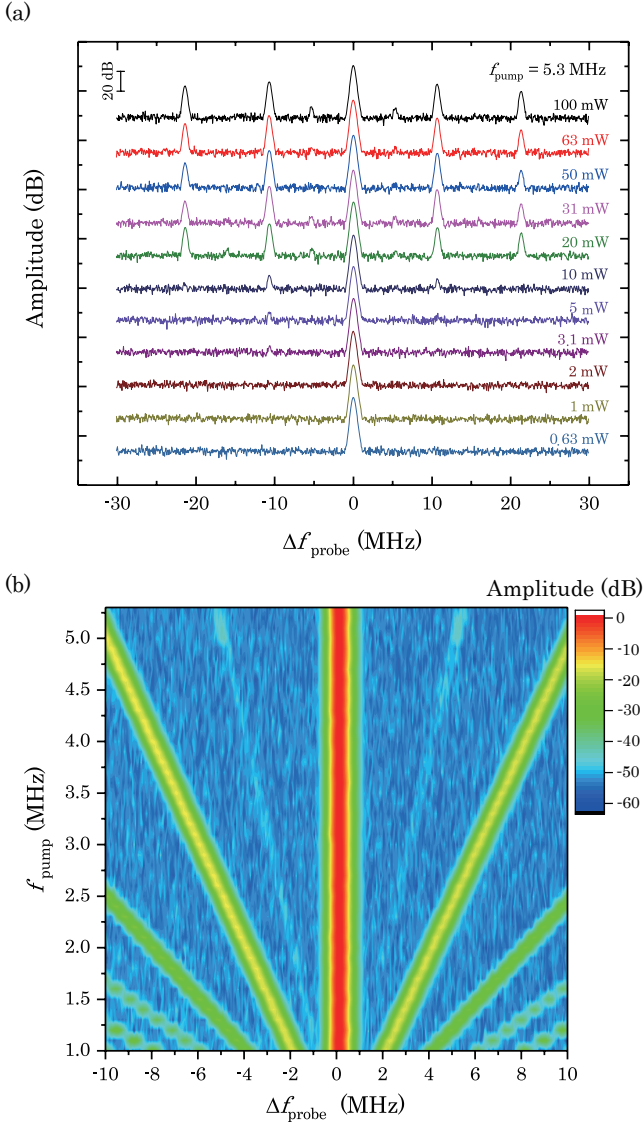}
	\caption{
	(color online) Characterization of the output frequencies of the pump field with a diplexer as functions of (a) the pump power with a fixed pump frequency of 5.3 MHz and (b) the pump frequency with a fixed pump power of 63 mW, with a probe frequency of 2822 MHz and probe power at 10 $\mu$W.
	}
	\label{figSpeAna}
\end{figure}

To estimate the output frequency characteristics of the diplexer, we measured its output as functions of the pump power and pump frequency while fixing the probe frequency at 2822 MHz and its power at 10 $\mu$W. We first measured the amplitude of the output frequency as a function of the pump power while fixing the pump frequency at 5.3 MHz. The results are shown in Fig.~\ref{figSpeAna}(a). We note that $\Delta f_\mathrm{probe}$ = 0 corresponds to the probe frequency and the scale bar in Fig.~\ref{figSpeAna}(a) indicates an amplitude difference of 20 dB. Figure~\ref{figSpeAna}(a) shows that there are several peaks when the pump power is over 10 mW, while there is only one peak, around 2822 MHz, when the pump power is below 5 mW. The peaks appear when the following conditions are satisfied: $f = f_\mathrm{probe} \pm n \times f_\mathrm{pump}$, where $f$, $f_\mathrm{probe}$, and $f_\mathrm{pump}$ are the peak frequency, probe frequency, and pump frequency, respectively, and $n$ is an integer. The peaks at $\left| n \right|$ = 1 have an amplitude of $\sim$ $-$48 dB and the peaks at $\left| n \right|$ = 2 have an amplitude of $\sim$ $-$15 dB when the pump power is 63 mW. We defined 0 dB as the amplitude of the peak at $n$ = 0. Next, we measured the amplitude of the output frequency as a function of pump frequency while fixing the pump power at 63 mW, as shown in Fig.~\ref{figSpeAna}(b). The figure shows that the peaks of $\left| n \right|$ = 1 have an amplitude of $\sim$ $-$50 dB and the peaks of $\left| n \right|$ = 2 have an amplitude of $\sim$ $-$15 dB. These results indicate that the expected effect on the ODMR measurements of the frequencies with $\left| n \right|$ = 2 is significantly higher than those with $\left| n \right|$ = 1. However, as discussed in the main text, the effect of the frequencies with $\left| n \right|$ = 2 is negligibly small in our main experiment.

\section{Nuclear magnetic resonance in NV center in diamond} \label{AppNV}
When the static magnetic field ($\bm{B}_0$) is applied to the NV axis parallel to the [111] direction of the diamond lattice, the spin Hamiltonian of the NV center~\cite{Doherty13} can be described as follows:
\begin{eqnarray}
\displaystyle{\cal{H}}_{NV} = D_{gs}S_z^2 + g_e\mu_BS_zB_0 + AS_zI_z + QI_z^2,
\end{eqnarray}
where $D_{gs}$, $A$, and $Q$ are the zero-field splitting parameter, the strength of the hyperfine interaction between the NV electron spin and $^{14}$N Nuclear spin, and the strength of the quadrupole interaction of $^{14}$N nuclear spins, respectively. $g_e$ and $\mu_B$ are the $g$-factor and Bohr magnetron of the NV electron spins, respectively. $S_z$ and $I_z$ are the electron and nuclear spins along the $z$ axis, which is parallel to the NV axis. We have neglected the anisotropic hyperfine term and nuclear Zeeman energy because they are negligibly small under $B_0 \sim$ 1.5 mT.

In our experiments, the NMR signals of the NV center can be observed via the NV electron spins. This means that the selection rules of ESR and NMR should be simultaneously satisfied. When circularly polarized mw and rf fields are applied to the $x$ and $y$ axes, the transition probability ($P$) is described by
\begin{equation}
P \propto B_1B_2 \bra{m_s, m_I}S_x\ket{m_s',m_I'}\bra{m_s', m_I'}I_y\ket{m_s'',m_I''},
\end{equation} 
where $B_1$ and $B_2$ are the strengths of the mw and rf fields, respectively. $m_s$ and $m_I$ are the electron and nuclear spins, respectively. Then, the following relation should be satisfied to observe the NMR signals of the NV center:
\begin{equation}
\bra{m_s, m_I}S_x\ket{m_s',m_I'}\bra{m_s', m_I'}I_y\ket{m_s'',m_I''} \ne 0.
\end{equation} 
This relation is satisfied when the following relations are satisfied:
\begin{eqnarray}
m_s -m_s' = \pm1, m_I-m_I' = 0, \nonumber \\
m_s' - m_s'' = 0, m_I'-m_I'' = \pm 1, \nonumber
\end{eqnarray}
Here, the observed NV electron spin resonances ($f_0$, $f_1$, and $f_{-1}$ )  due to the hyperfine interaction with $^{14}$N nuclear spin are described by
\begin{eqnarray}
f_0 &=& D_{gs}-g_e\mu_BB_0, \\
f_1 &=& D_{gs}-g_e\mu_BB_0 + A, \\
f_{-1} &=& D_{gs}-g_e\mu_BB_0 - A, 
\end{eqnarray}
when the NV electron spin is $m_s= -1$. Then, the observed NMR signals of the $^{14}$N nuclear spin in the NV center are described by
\begin{eqnarray}
f_1 &=& E_{0,0} - E_{-1,+1} = f_0 - Q - A, \nonumber \\
f_2 &=& E_{0,0} - E_{-1,-1} = f_0 - Q + A, \nonumber \\
f_3 &=& E_{0,1} - E_{-1,0} =f_1 + Q -A, \nonumber \\
f_4 &=& E_{0,-1}-E_{-1,0} = f_{-1} +Q+A. \nonumber
\end{eqnarray}
These equations indicate that the NMR signals can be observed when the sum and difference of the irradiated mw and rf frequencies correspond to $f_1$, $f_2$, $f_3$, or $f_4$. 

\section{Multi-photon resonances in NV center} \label{AppMPNV}
\begin{figure}
	\includegraphics[width=8cm,clip]{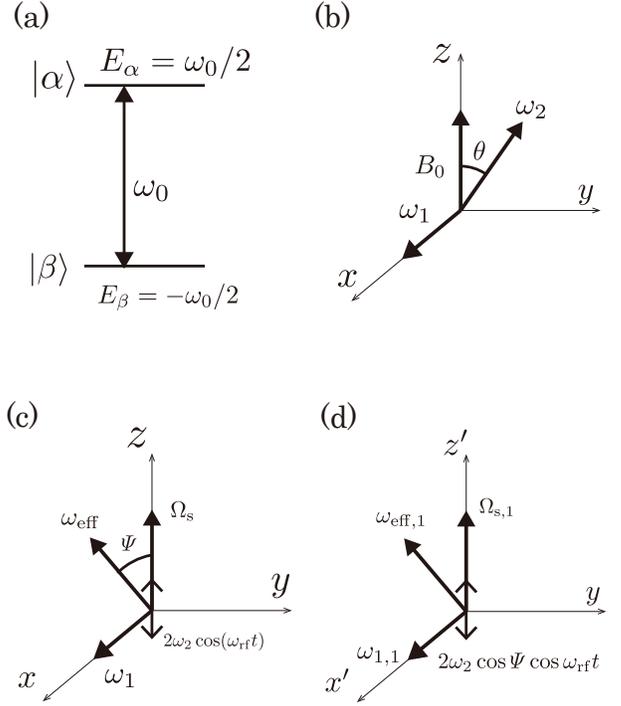}
	\caption{
	(a) Two-level system of$E_\alpha$ and $E_\beta$. (b) Configuration of a static magnetic field ($B_0$), mw field ($\omega_1$), and rf field ($\omega_2$) in the laboratory frame. (c) Configuration of the singly rotating frame. (d) Configuration of the doubly rotating frame.
	}
	\label{figMPResonance}
\end{figure}

In this study we used mw and rf fields to perform multiphoton resonances in a single NV center. To explain our results, we consider the configuration in the laboratory frame depicted in Fig.~\ref{figMPResonance}(a). Figure~\ref{figMPResonance}(a) shows a two-level system ($\ket{\alpha}$ and $\ket{\beta}$) with energies $E_\alpha = \omega_0/2$ and $E_\beta = -\omega_0/2$ under the irradiation of mw and rf fields~\cite{BurgetJHB61,GramoveJMR00}. In the case of an NV center, these two states correspond to, for example, $\left| 0,0 \right>$ and $ \left| -1, 0 \right>$. A static magnetic field ($\bm{B}_0$) and an mw (amplitude 2$\omega_1$ and frequency $\omega_\mathrm{mw}$) field are applied in the $z$ and $x$ directions, respectively. An rf (amplitude 2$\omega_2$ and frequency $\omega_\mathrm{rf}$) field was applied with an angle $\theta$, which corresponds to the angle between the $\bm{B}_0$ and rf field depicted in Fig.~\ref{figMPResonance}(b). When the rf field is oriented along the $z$ direction ($\theta$=0), the Hamiltonian ($\displaystyle{\cal{H}}$) is described by 
\begin{equation}
\displaystyle{\cal{H}} = \omega_0 S_z + 2\omega_1\cos{\left(\omega_\mathrm{mw}t\right)}S_x + 2\omega_2\cos{\left(\omega_\mathrm{rf}t\right)}S_z,
\end{equation} 
where $\omega_0$ is the resonant frequency of the two-level system. $S_z$ and $S_x$ are defined as the NV electron spins along the $z$ and $x$ directions, respectively. Then, $\displaystyle{\cal{H}}$ can be transformed to the rotating frame at $\omega_\mathrm{mw}$ described by 
\begin{equation}
\displaystyle{\cal{H'}}_{s} = \Omega_\mathrm{s} S_z + \omega_1S_x + 2\omega_2 \cos{\left(\omega_\mathrm{rf}t\right)}S_z,
\label{eqSRFM}
\end{equation} 
where $\Omega_\mathrm{s}$ is the resonance offset of $\omega_0 -\omega_\mathrm{mw}$, and the configuration of the singly rotating frame is depicted in Fig.~\ref{figMPResonance}(c). Figure~\ref{figMPResonance}(c) shows that the effective nutation frequency, $\omega_\mathrm{eff}$, is tilted from $z$ by $\varPsi$, which is defined as $\tan^{-1}\left(\omega_1/\Omega_\mathrm{s}\right)$. Then, this frame is transformed to a rotating frame with $\omega_\mathrm{rf}$, as depicted in Fig.~\ref{figMPResonance}(d)~\cite{BoscainoPhys86,KalinJMR03}. When the counter-rotating term is neglected, the Hamiltonian of the doubly rotating frame is described by 
\begin{equation}
\displaystyle{\cal{H'}}_{s} = \Omega_\mathrm{s,1} S_z + \omega_{1,1} S_x,
\label{eqSRM}
\end{equation} 
where the resonance offset,  $\Omega_\mathrm{s,1}$, is defined by 
\begin{equation}
\Omega_\mathrm{s,1} = \omega_\mathrm{eff}- \omega_{rf} \approx \Omega_s-\omega_\mathrm{rf},
\end{equation}
for $\omega_1 \ll \Omega_s$ because the mw field is considered to be a perturbed field. Also, the effective field amplitude $\omega_{1,1}$ is described by 
\begin{equation}
\omega_{1,1} = -\omega_2\sin{\varPsi} = -\frac{\omega_1\omega_2}{\sqrt{\omega_1^2 + \Omega_\mathrm{s}^2}} = -\frac{\omega_1\omega_2}{\Omega_\mathrm{s,1}+\omega_\mathrm{rf}}.
\end{equation}
When we consider a near resonance condition $\left(\Omega_{s,1} \ll \omega_\mathrm{rf}\right)$, $\omega_{1,1} $ can be described as 
\begin{equation}
\omega_{1,1} = -\frac{\omega_1\omega_2}{\omega_\mathrm{rf}}.
\end{equation}
Then, $\displaystyle{\cal{H'}}_{d}$ is described by
\begin{equation}
\displaystyle{\cal{H'}}_{d} =\left( \Omega_s-\omega_\mathrm{rf}\right)S_z - \frac{\omega_1\omega_2}{\omega_\mathrm{rf}}S_x. 
\label{eqDRM}
\end{equation}
This equation shows that two-photon transitions occur when $\omega_\mathrm{s} = \omega_\mathrm{mw} + \omega_\mathrm{rf}$ is satisfied. In addition, the above discussion can be extended for multiple-photon resonances ($\sigma + k \times \pi$, where $k$ is integer) in the toggling frame (see the details discussed in Ref.~\onlinecite{KalinJMR03}). The first-order toggling Hamiltonian is described by 
\begin{equation}
\displaystyle{\cal{H'}}_{t} = \left( \Omega_s-k\omega_\mathrm{rf}\right)S_z + \omega_1J_{-k}\left(\frac{2\omega_2}{\omega_\mathrm{rf}}\right)S_x, 
\label{eqC8}
\end{equation}
where $J_n(z)$ is the Bessel function of the first kind and $n$ is an integer. If $2\omega_2/\omega_\mathrm{rf} \ll 1$ is satisfied, the Bessel function can be described by 
\begin{equation}
J_n\left(\frac{2\omega_2}{\omega_\mathrm{rf}}\right) \approx \frac{1}{n!}\left(\frac{2\omega_2}{\omega_\mathrm{rf}}\right)^n.
\end{equation}
Then, the effective transition amplitude is described by 
\begin{equation}
\omega_{1,k} \approx \omega_1\frac{[\mathrm{sign}(-k)]^k}{|k|!}\left(\frac{\omega_2}{\omega_\mathrm{rf}}\right). 
\end{equation}

\section{Absorption signals of two-photon magnetic resonance (TPMR)} \label{AppTPMR}
The magnetization vector $\bm{M = (M_x, M_y, M_z)}$ under static and oscillating magnetic fields is described by the Bloch equation. The Bloch equation of Eq. (\ref{eqSRM}) is described by
\begin{eqnarray}
\frac{d}{dt}\bm{\sigma} &=& -i 
\begin{pmatrix}
&  -\Omega_\mathrm{s}&  \\
\Omega_\mathrm{s} &  & \omega_1 \\
& -\omega_1&
\end{pmatrix}\bm{\sigma}  \nonumber \\
&& -\begin{pmatrix}
1/T_2 & &  \\
& 1/T_2 &  \\
& & 1/T_1
\end{pmatrix}(\bm{\sigma} - \bm{\sigma}_0), 
\end{eqnarray}
where $\bm{\sigma}$ is a density operator and  $\bm{\sigma}_0$ is the density operator for $(0,0,-1)$ at thermal equilibrium~\cite{KalinJMR03}. $\bm{\sigma}$ has the following relation with the magnetization, $M_q = 2 tr\left\{\sigma S_q\right\}M_0$ with $q = x, y ,z$.  $T_1$ and $T_2$ are the longitudinal and transverse relaxation times, respectively. The observed absorption signal is given by the steady state solution of $\frac{d}{dt}\bm{\sigma} =0$. As a result, the observed absorption signal is described by
\begin{equation}
<S_y>  =   -\frac{\omega_1T_2}{\left(1+\omega_1^2 T_1T_2\right)+\Omega_\mathrm{s}^2T_2^2}.  
\end{equation}

\begin{figure}
	\includegraphics[width=8cm,clip]{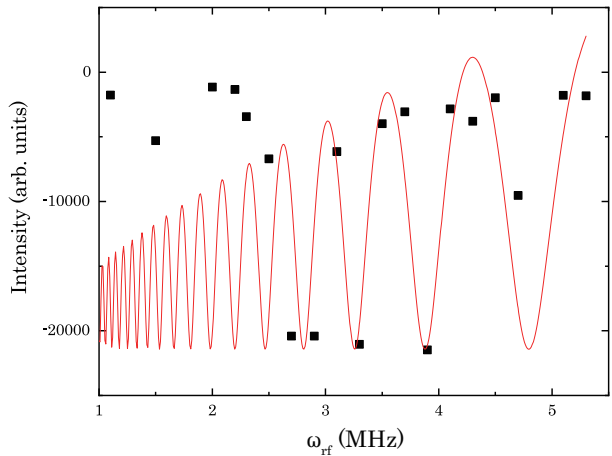}
	\caption{
	Intensity of the TPMR as a function of rf frequency. The data points show the experimental results and the solid line shows the simulation result (details in main text). 
	}
	\label{figMPRExVTheo}
\end{figure}

Next, we consider the absorption signals of the TPMRs. The effective Hamiltonian of the doubly rotating frame is shown in Eq. (\ref{eqDRM}). $\bm{\sigma}_0$ is transformed in the doubly rotating frame to $\bm{\sigma}_d(t) = R(t)\bm{\sigma}_0R^{-1}(t)$, where $R(t)$ is a rotation operator defined as 
\begin{equation}
R(t) = \exp{\left\{i\left[k\omega_\mathrm{rf}t + \left(\frac{2\omega_2}{\omega_\mathrm{rf}}\right)\sin{\left(\omega_\mathrm{rf}t\right)}\right]S_z\right\}}
\end{equation}
Then, the absorption signals of the two-photon and single-photon magnetic resonances are described by
\begin{equation}
<S_y>  = \sum_{k=-1}^{1} \frac{\omega_1T_2J_{k}\left(\frac{2\omega_2}{\omega_\mathrm{rf}}\right)^2
}{1+\omega_1^2J_{k}\left(\frac{2\omega_2}{\omega_\mathrm{rf}}\right)^2T_1 T_2 + \left(\Omega_s-k\omega_\mathrm{rf}\right)^2T_2^2}. 
\label{EqTPMRAmp}
\end{equation}
Finally, we compared the theory of Eq. (\ref{EqTPMRAmp}) with the experimental results. The data points and solid line in Fig.~\ref{figMPRExVTheo} show the experimental results and results of curve fitting using Eq. (\ref{EqTPMRAmp}), respectively. To reduce uncertainties in the curve fitting, we fixed $\omega_1$ = 90 kHz from our experimental conditions and used typical values of $T_1$ = 6 msec~\cite{NeumannScience08} and $T_2$ = 1 $\mu$sec~\cite{MizuochiPRB09}. While the results for $\omega_\mathrm{rf} > $ 2.5 MHz agree with Eq. (\ref{EqTPMRAmp}), the results for $\omega_\mathrm{rf} < $ 2.5 MHz do not agree with the theory. We need to consider the effect of the $^{14}$N nuclear spins. In our experimental conditions, the $^{14}$N nuclear spins are not polarized and the hyperfine interaction between the NV electron spin and $^{14}$N nuclear spin complicate the spectrum complex. Thus, it is hard to precisely explain the result for $\omega_\mathrm{rf} <$ 2.5 using only the simple theory of Eq. (\ref{EqTPMRAmp}). This affect is also experimentally observed and explained in Ref.~\onlinecite{ChildressPRA10}.

%

\end{document}